\documentclass[12pt]{elsart}
\usepackage{epsfig}

\begin{document}
\vspace{-1.0cm}
\begin{flushleft}
{\normalsize MIT-CTP-3001} \hfill\\
{\normalsize September 2000} 
\end{flushleft}
\vspace{1.0cm}
\begin{frontmatter}
  \title{\bf Perturbative renormalization of moments of quark momentum,
   helicity and transversity distributions with overlap and Wilson fermions}
  
   \author{Stefano Capitani}
  
   \address{Center for Theoretical Physics, Laboratory for Nuclear 
   Science \\ Massachusetts Institute of Technology \\ 77 Massachusetts Ave., 
   Cambridge, MA 02139, USA \\ \ \\ {\sf E-mail: stefano@mitlns.mit.edu} }

  \date{ }

\begin{abstract}

Using overlap as well as Wilson fermions, we have computed the one-loop 
renormalization factors of ten non-singlet operators which measure the 
third moment of quark momentum and helicity distributions (the lowest two 
having been computed in a previous paper), as well as the lowest three 
moments of the $g_2$ structure function and the lowest two non-trivial 
moments of the $h_1$ transversity structure function (plus the tensor charge). 
These factors are needed to extract physical observables from Monte Carlo 
simulations of the corresponding matrix elements.

An exact chiral symmetry is maintained in our calculations with overlap 
fermions, and its most important consequence here is that the operators 
measuring $g_2$ do not show any of the power-divergent mixings with 
operators of lower dimension which are present in the Wilson case. 
Many of our results for Wilson fermions are also new; for the remaining 
ones, we agree with the literature except in one case.
The computations have been carried out using the symbolic language FORM, 
in a general covariant gauge, which turns out also to be useful in checking 
the gauge-invariance of the final results.

\vspace{0.5 cm}

{\bf Keywords:} Lattice QCD, Overlap-Dirac operator, Chirality,
                Structure functions, Lattice perturbation theory,
                Lattice renormalization.

{\bf PACS numbers:} 11.15.-q, 11.15.Ha, 12.38.Gc, 13.60.Hb.

\end{abstract}
\end{frontmatter}

\newpage

\section{Introduction}

Overlap fermions~\cite{overlap} have emerged in recent years as one of the 
most promising formulations for simulating on the lattice theories that 
possess an exact chiral symmetry. The overlap-Dirac operator proposed by 
Neuberger~\cite{n2} is one of the solutions of the Ginsparg-Wilson 
relation~\cite{gw}
\begin{equation}
\gamma_5 D + D \gamma_5 = a \frac{1}{\rho} D \gamma_5 D ,
\label{eq:gw}
\end{equation}
and allows the realization of an exact chiral symmetry also at non-zero 
lattice spacing~\cite{l2} without giving up other important symmetries. 
Although simulations with the Neuberger operator look computationally 
demanding when compared with Wilson fermions, progress is under way 
in simulating overlap fermions and in improving the efficiencies of their 
simulation algorithms~\cite{hjl,sim,lat00g,lat00l}. 
Results of some recent simulations show remarkable evidence of the
signatures of the exact chiral symmetry. For example, Ref.~\cite{lat00g} 
provides strong numerical evidence that the pion mass approaches zero
in the limit $m_0a \rightarrow 0$ without additive renormalization, 
in the two-dimensional Schwinger model, and progress is on the way for QCD; 
in Ref.~\cite{lat00l} it is shown how the Gell-Mann-Oakes-Renner relation, 
which is a test of chiral symmetry at finite lattice spacing, is satisfied 
to better than 1\% down to quark masses as small as $m_0a = 0.006$. 
There is some confidence that overlap fermions will at the end turn out to 
be not too computationally demanding, and in any case one has to take into
account all the advantages of having an exact chiral symmetry as well as the 
rapid progress in computer technology that is taking place these days.

In this paper we present the computation of the renormalization factors of 
many operators that measure moments of various structure functions, using 
overlap fermions. The operators that we consider include all three parton 
distributions that characterize the quarks in the nucleon and provide a 
complete description of quark momentum and spin at leading twist: the momentum
distribution $q(x,Q^2)$, the helicity distribution $\Delta q(x,Q^2)$ and the 
transversity distribution $\delta q(x,Q^2)$. We also study the $g_2$ structure
function, which receives contributions from twist-3 operators. For each one
of these structure functions, we have included in our study all the moments 
that can be measured by operators for which all tensor indices are distinct.
On the lattice with overlap fermions, all these operators are then 
multiplicatively renormalized.

One of the advantages of overlap fermions is that lattice renormalization 
does not induce any mixings with operators of the wrong chirality, as instead 
it happens with Wilson fermions. While this issue is of the utmost importance 
in lattice simulations concerning weak decays and quantities like the 
CP-violation parameter $\epsilon'/\epsilon$~\cite{cg}, it is also relevant 
here for some of the operators that measure the $g_2$ structure function, 
which for Wilson fermions mix with coefficients diverging as $1/a$, while 
for overlap fermions they are multiplicatively renormalized (see Sect.~5). 
Another advantage of overlap fermions is the reduction of the number of 
independent renormalization factors in a given physical situation, as can 
be seen here with the operators measuring unpolarized and polarized structure 
functions which differ by a $\gamma_5$ matrix.
Significant computational gains are achieved in the improvement of the action 
(which is not needed at all) and of the operators, as the construction of 
improved operators and the calculation of their renormalization constants is 
much simpler for overlap than for Wilson fermions (see Sect.~4). We think that
this is really an important point for the operators considered in this paper, 
as the full $O(a)$ improvement of DIS operators with Wilson fermions involves 
a significant amount of careful and cumbersome calculations. 

Among the other solutions of the Ginsparg-Wilson relation, a popular one 
is given by domain-wall fermions, where however the decoupling of the chiral 
modes is achieved only in the limit in which the number of points in the fifth
additional dimension goes to infinity. One of the most attractive features 
of overlap fermions is that, on the contrary, chiral symmetry is fully 
preserved for any finite volume of the lattice.
As the authors of Ref.~\cite{hjl2} state, ``in practical applications, it is 
our present experience that it is easier to control chiral symmetry violations
with Neuberger's operator''.

The renormalization with overlap fermions of the lowest two moments of the 
quark momentum and helicity distributions has been already carried out in 
Ref.~\cite{primidue}, and we refer to that paper for the general framework 
and conventions, as well as for the Feynman rules of the overlap theory that 
we use here. We recall only that the explicit form of the overlap-Dirac 
operator is 
\begin{equation}
D_N = \frac{1}{a} \rho \, \Big[ 1+ \frac{X}{\sqrt{X^\dagger X}} \Big] ,
\label{eq:dn}
\end{equation}
where
\begin{equation}
X=D_W -\frac{1}{a} \rho ,
\end{equation}
with  $0 <\rho <2r$, in terms of the usual Wilson-Dirac operator
\begin{equation}
D_W = \frac{1}{2} \Big[ \gamma_\mu (\nabla^\star_\mu + \nabla_\mu)
- a r \nabla^\star_\mu \nabla_\mu \Big] ,
\end{equation}
\begin{equation}
\nabla_\mu \psi (x) = 
\frac{1}{a} \Big[ U(x,\mu)\psi(x+a\hat{\mu}) - \psi (x) \Big] .
\end{equation}

Since the renormalization constants for many of the operators considered 
in this paper had not even been computed with Wilson fermions, we have 
calculated them for all the operators considered here with Wilson fermions 
too, and in this way we have also checked some old results in the literature, 
finding a discrepancy in one case.

This paper is organized as follows: in Sect.~2 we introduce the various 
operators of which we have computed the renormalization constants, in Sect.~3 
their renormalization on the lattice is discussed and some details about the 
perturbative calculations are given, in Sect.~4 we discuss the advantages of 
using overlap over Wilson fermions with regard to the improvement, and in 
Sect.~5 we present the complete results. In the appendices we give the results
for the quark self-energy, both for overlap and for Wilson fermions, and for 
the individual proper diagrams.

\section{Moments of structure functions}

We have considered in this paper the lowest moments of a few structure 
functions which give a complete description of quark momentum and spin
at leading twist: the momentum distribution $q(x,Q^2)$, measured by the 
$F_1$ and $F_2$ unpolarized structure functions, the helicity distribution 
$\Delta q(x,Q^2)$, measured by the $g_1$ structure function, and the 
transversity distribution $\delta q(x,Q^2)$, measured by the $h_1$ 
structure function~\cite{transversity,am}. We have also considered the $g_2$
structure function~\cite{jcjj}, which measures the transverse spin, is chiral 
even and contains at leading order a twist-3 piece; it is one of the most 
accessible higher-twist quantities. The $h_1$ structure function is instead 
chiral odd and as such does not arise in inclusive DIS and can be measured 
instead in polarized Drell-Yan processes. 
For some more detailed discussions of deep inelastic scattering on the 
lattice, see Refs.~\cite{cr,bbcr,gea,gea2,calcoli,petronzio,primidue} 
and references therein.

The moments of the various distributions are, for a given flavor~\footnote{We 
have chosen our conventions in such a way that there is a manifest symmetry 
between $v_n$ and $a_n$. The operators corresponding to $v_n$ and $a_n$ have 
the same number of covariant derivatives, corresponding chiral properties, 
and the same renormalization constant for overlap fermions~\cite{primidue}, 
as well as obviously in any continuum scheme. The same convention was also 
used in Ref.~\cite{primidue} and can be found for example in~\cite{manohar}, 
but in other papers like e. g.~\cite{gea,calcoli} $a_n$ has a correspondence
with $v_{n+1}$.

In all operators considered here the subscript $n$ equals the number of 
symmetrized Lorentz indices. This means that for the operators measuring 
$d_n$ our convention is still the same as~\cite{gea,calcoli}.},
\begin{eqnarray}
2 \int_0^1 dx \, x^{n-1} F_1 (x,Q^2) &=& C_{1,n} 
\Big( \frac{Q^2}{\mu^2}, g(\mu) \Big) \, v_n (\mu) \nonumber \\
\int_0^1 dx \, x^{n-2} F_2 (x,Q^2) &=& C_{2,n} 
\Big( \frac{Q^2}{\mu^2}, g(\mu) \Big) \, v_n (\mu) \label{eq:momsf} \\
2 \int_0^1 dx \, x^{n-1} g_1 (x,Q^2) &=& \frac{1}{2} E_{1,n} 
\Big( \frac{Q^2}{\mu^2}, g(\mu) \Big) \, a_n (\mu) \nonumber \\
2 \int_0^1 dx \, x^{n-1} g_2 (x,Q^2) &=& \frac{1}{2} \frac{n-1}{n}  
\Bigg( E_{2,n} \Big( \frac{Q^2}{\mu^2}, g(\mu) \Big) \, d_{n-1} 
(\mu) \nonumber \\
& & \qquad \qquad \qquad
-E_{1,n} \Big( \frac{Q^2}{\mu^2}, g(\mu) \Big) \, a_n (\mu) 
\Bigg) \nonumber \\
2 \int_0^1 dx \, x^{n-1} h_1 (x,Q^2) &=& \frac{1}{2} B_{1,n} 
\Big( \frac{Q^2}{\mu^2}, g(\mu) \Big) \, t_n (\mu) , \nonumber 
\end{eqnarray}
where $C_{i,n}$, $E_{i,n}$ and $B_{1,n}$ denote the appropriate Wilson 
coefficients in the OPE expansions, which can be computed in continuum 
perturbation theory. The moments of the helicity and transversity 
distributions are given by the formulae
\begin{eqnarray}
a_{n+1} &=& 2 \Delta^{(n)} q , \qquad \Delta^{(n)} q (\mu) =  
\int_0^1 dx \, x^n \Delta q (x,\mu) \\
t_{n+1} &=& 2 \delta^{(n)} q , \qquad \delta^{(n)} q  (\mu) =  
\int_0^1 dx \, x^n \delta q (x,\mu) ,
\nonumber
\end{eqnarray}
and the axial charge is $\Delta^{(0)} u - \Delta^{(0)} d = g_A = 1.26 $.

The quark operators that correspond to the moments are~\footnote{We use 
$\stackrel{\phantom{\rightarrow}}{D}=\stackrel{\rightarrow}{D}
-\stackrel{\leftarrow}{D}$, with the following lattice discretizations:
\begin{eqnarray}
\stackrel{\rightarrow}{D_\mu} \psi (x)  &=& \frac{1}{2a} \Big[U (x,\mu) 
\psi(x+a\hat{\mu}) -U^\dagger (x-a\hat{\mu},\mu) \psi(x-a\hat{\mu}) \Big] \\
\bar{\psi} (x) \stackrel{\leftarrow}{D_\mu} &=& \frac{1}{2a} \Big[
\bar{\psi} (x+a\hat{\mu}) U^\dagger (x,\mu) - \bar{\psi}(x-a\hat{\mu}) 
U(x-a\hat{\mu},\mu) \Big] . \nonumber
\end{eqnarray}}
\begin{eqnarray}
O_{\mu_1 \cdots \mu_n} &=& \Bigg( \frac{\rm i}{2} \Bigg)^{n-1} 
\bar{\psi} \gamma_{\mu_1} D_{\mu_2} \cdots D_{\mu_n} \psi -\mbox{traces} 
\nonumber \\
O^5_{\mu_1 \cdots \mu_n} &=& \Bigg( \frac{\rm i}{2} \Bigg)^{n-1}
\bar{\psi} \gamma_{\mu_1} \gamma_5 D_{\mu_2} \cdots D_{\mu_n} \psi 
-\mbox{traces} \\
O^h_{\mu_1 \cdots \mu_n} &=& \Bigg( \frac{\rm i}{2} \Bigg)^{n-2} 
\bar{\psi} \sigma_{\mu_1\mu_2} \gamma_5 D_{\mu_3} \cdots D_{\mu_n} \psi 
-\mbox{traces} . \nonumber
\end{eqnarray}
The matrix elements of the above operators on polarized quark states are
\begin{eqnarray}
\frac{1}{2}\sum_s \langle \vec{p},\vec{s} | O_{\{ \mu_1 \cdots
\mu_n \} } | \vec{p}, \vec{s} \rangle &=& 2 \, v_n \, \Big[ p_{\mu_1}
\cdots p_{\mu_n} -\mbox{traces} \Big] \nonumber \\
\langle \vec{p}, \vec{s} | O^5_{\{ \mu_1 \cdots \mu_n \}} |
\vec{p}, \vec{s} \rangle &=& \frac{1}{n} \, a_n \, \Big[ 
s_{\{\mu_1} p_{\mu_2} \cdots p_{\mu_n\}} +\cdots -\mbox{traces} \Big] \\
\langle \vec{p}, \vec{s} | O^5_{[\mu_1 \{ \mu_2 ] \cdots \mu_n
\} } | \vec{p}, \vec{s} \rangle &=& \frac{1}{n} \, d_{n-1} \, \Big[ 
s_{[\mu_1} p_{\{\mu_2]} p_{\mu_3} \cdots p_{\mu_n\}} +\cdots -\mbox{traces} 
\Big] \nonumber \\
\langle \vec{p}, \vec{s} | O^h_{\mu_1 \{ \mu_2 \cdots \mu_n \}} |
\vec{p}, \vec{s} \rangle &=& \frac{1}{m_N} \, t_{n-1} \, \Big[ 
s_{[\mu_1} p_{\{\mu_2]} p_{\mu_3} \cdots p_{\mu_n\}} +\cdots -\mbox{traces} 
\Big] , \nonumber 
\end{eqnarray}
where $s_\mu$ is the polarization vector of the nucleon, with $s^2=-m_N^2$.

The moments of the various distributions can be studied from first principles
by performing lattice Monte Carlo simulations of the above matrix
elements, which determine then the various $v_n$, $a_n$, $d_n$ and $t_n$ 
quantities. These bare numbers need however to be renormalized. Under 
renormalization, the quark operators corresponding to momentum and helicity 
distributions mix in the flavor singlet case with operators that measure the 
corresponding gluon distributions. We consider in this paper only flavor 
non-singlet operators, so that the mixing with gluon operators is forbidden. 
The operators corresponding to the transversity distribution however do not 
have any gluon mixing even in the flavor-singlet case, as there is no gluonic 
transversity at leading twist, i. e. no chiral-odd gluon operator can be 
constructed. 

In a previous paper we have calculated the renormalization factors of a few 
operators which measure $v_2$, $v_3$, $a_2$ and $a_3$~\cite{primidue}; here 
we compute the renormalization constants of the operators 
\begin{eqnarray}
O_{v_4,d} &=& \bar{\psi} \gamma_{\{4} D_1 D_2 D_{3\}} \psi \\
O_{v_4,e} &=& \bar{\psi} \gamma_{\{4} D_4 D_1 D_{1\}} \psi
 +\bar{\psi} \gamma_{\{3} D_3 D_2 D_{2\}} \psi \nonumber \\
 && -\bar{\psi} \gamma_{\{4} D_4 D_2 D_{2\}} \psi
    -\bar{\psi} \gamma_{\{3} D_3 D_1 D_{1\}} \psi ,
\end{eqnarray}
which measure the third moment of the quark momentum distribution, and
\begin{eqnarray}
O_{a_4,d} &=& \bar{\psi} \gamma_{\{4} \gamma_5 D_1 D_2 D_{3\}} \psi \\
O_{a_4,e} &=& \bar{\psi} \gamma_{\{4} \gamma_5 D_4 D_1 D_{1\}} \psi
 +\bar{\psi} \gamma_{\{3} \gamma_5 D_3 D_2 D_{2\}} \psi \nonumber \\
 && -\bar{\psi} \gamma_{\{4} \gamma_5 D_4 D_2 D_{2\}} \psi
    -\bar{\psi} \gamma_{\{3} \gamma_5 D_3 D_1 D_{1\}} \psi ,
\end{eqnarray}
which measure the third moment of the $g_1$ quark helicity distribution. 
Together with the computations in Ref.~\cite{primidue}, they complete the 
calculation of the renormalization factors of the lowest three moments of 
these distributions; each continuum operator has been considered twice on 
the lattice by choosing its indices in two different ways, corresponding to 
two different representations of the hypercubic group for the same moment, 
which renormalize independently on the lattice.

We have computed also the renormalization constants of the operators
\begin{eqnarray}
O_{d_1} &=& \bar{\psi} \gamma_{[4} \gamma_5 D_{1]} \psi \\
O_{d_2} &=& \bar{\psi} \gamma_{[4} \gamma_5 D_{\{1]} D_{2\}} \psi \\
        &=& \frac{1}{2} \Big( \bar{\psi} \gamma_4 \gamma_5 D_1 D_2 \psi 
               +\bar{\psi} \gamma_4 \gamma_5 D_2 D_1 \psi 
               -\bar{\psi} \gamma_1 \gamma_5 D_4 D_2 \psi 
               -\bar{\psi} \gamma_1 \gamma_5 D_2 D_4 \psi \Big) \nonumber \\
O_{d_3} &=& \bar{\psi} \gamma_{[4} \gamma_5 D_{\{1]} D_2 D_{3\}} \psi \\
        &=& \frac{1}{6} \Big( \bar{\psi} \gamma_4 \gamma_5 D_1 D_2 D_3 \psi
              +\bar{\psi} \gamma_4 \gamma_5 D_1 D_3 D_2 \psi 
              +\bar{\psi} \gamma_4 \gamma_5 D_2 D_1 D_3 \psi \nonumber \\
        && \quad  +\bar{\psi} \gamma_4 \gamma_5 D_2 D_3 D_1 \psi 
              +\bar{\psi} \gamma_4 \gamma_5 D_3 D_1 D_2 \psi
              +\bar{\psi} \gamma_4 \gamma_5 D_3 D_2 D_1 \psi \nonumber \\
        && \quad  -\bar{\psi} \gamma_1 \gamma_5 D_4 D_2 D_3 \psi
              -\bar{\psi} \gamma_1 \gamma_5 D_4 D_3 D_2 \psi 
              -\bar{\psi} \gamma_1 \gamma_5 D_2 D_4 D_3 \psi \nonumber \\
        && \quad -\bar{\psi} \gamma_1 \gamma_5 D_2 D_3 D_4 \psi 
              -\bar{\psi} \gamma_1 \gamma_5 D_3 D_4 D_2 \psi
              -\bar{\psi} \gamma_1 \gamma_5 D_3 D_2 D_4 \psi \Big) , \nonumber
\end{eqnarray}
which taken together with $O_{a_2}$, $O_{a_3}$ and $O_{a_4}$ determine the 
lowest three moments of the $g_2$ structure function in Eq.~(\ref{eq:momsf}). 
In the Wilson case each one of the $O_{d_n}$ operators,
\begin{equation} 
\bar{\psi} \gamma_{[\sigma } \gamma_5 D_{\{\mu_1]} D_{\mu_2} \cdots 
D_{\mu_n\}} \psi ,
\end{equation} 
mixes, due to the breaking of chirality, with a lower-dimensional operator 
which in the continuum OPE has a mass coefficient~\cite{jcjj},
\begin{equation} 
m_q \, \bar{\psi} \gamma_{[\sigma } \gamma_5 \gamma_{\{\mu_1]} D_{\mu_2} 
\cdots D_{\mu_n\}} \psi \label{eq:massope},
\end{equation} 
but on the lattice this mass becomes a $1/a$ divergent coefficient. 
This mixing is forbidden in the overlap case, and the $O_{d_n}$ operators 
are multiplicatively renormalized. Thus the overlap makes a complete 
perturbative renormalization of these operators possible. 

Regarding the transversity distribution $h_1$, we have considered the 
twist-two operators that measure the tensor charge and the lowest two 
non-trivial moments,
\begin{eqnarray}
O_{t_1} &=& \bar{\psi} \sigma_{41} \gamma_5 \psi \\
O_{t_2} &=& \bar{\psi} \sigma_{4\{1} \gamma_5 D_{2\}} \psi \\
O_{t_3} &=& \bar{\psi} \sigma_{4\{1} \gamma_5 D_2 D_{3\}} \psi .
\end{eqnarray}

The moments of the various distributions shown above are all the ones 
that can be associated with an operator which has all the indices different 
from each other. There is then no mixing due to the breaking of the 
Lorentz group to the hypercubic group. In fact, all the above operators 
are multiplicatively renormalized in the overlap case; this is true also 
for the particular combinations in $O_{v_4,e}$ and $O_{v_4,e}$ in which 
some of the indices are equal. 
The $O_{d_n}$ operators are not multiplicatively renormalized when using 
Wilson fermions, but their mixings are in this case due only to the breaking 
of chiral symmetry. 

All operators measuring higher moments of the above distributions necessarily 
have at least two equal indices. In fact, the operators which would be next 
in the ladder of moments have five Lorentz indices and therefore two of them 
have to be equal:
\begin{eqnarray}
O_{v_5} &=& \bar{\psi} \gamma_{\{4} D_1 D_1 D_2 D_{3\}} \psi \nonumber \\
O_{a_5} &=& \bar{\psi} \gamma_{\{4} \gamma_5 D_1 D_1 D_2 D_{3\}} \psi \\
O_{d_4} &=& \bar{\psi} \gamma_{[4} \gamma_5 D_{\{1]} D_1 D_2 D_{3\}} \psi
\nonumber \\
O_{t_4} &=& \bar{\psi} \sigma_{4\{1} \gamma_5 D_1 D_2 D_{3\}} \psi . \nonumber 
\end{eqnarray} 
This leaves open the possibility of mixing with same-dimension or 
(more catastrophic) lower-dimension operators. We do not address here
this problem, which is left for further studies.

\section{Perturbative renormalization}

The raw numbers extracted from Monte Carlo simulations need to be renormalized
to physical continuum quantities. The connection is given by 
\begin{equation}
\langle O_i^{\rm cont} \rangle = \sum_j \Bigg( \delta_{ij} 
-\frac{g_0^2}{16 \pi^2} \Big( R_{ij}^{\rm lat} -R_{ij}^{\rm cont} \Big) \Bigg) 
\cdot \langle O_j^{\rm lat} \rangle ,
\label{eq:matching}
\end{equation}
where
\begin{equation}
\langle O_i^{\rm cont,lat} \rangle = \sum_j \Bigg( \delta_{ij} 
+ \frac{g_0^2}{16 \pi^2} R_{ij}^{\rm cont,lat} \Bigg) 
\cdot \langle O_j^{\rm tree} \rangle
\end{equation}
are the continuum and lattice 1--loop expressions respectively, and the 
tree-level matrix element is the same in both cases. Their difference 
$\Delta R_{ij} = R_{ij}^{\rm lat} - R_{ij}^{\rm cont}$ enters then in the 
gauge-invariant renormalization factors
\begin{equation}
Z_{ij} (a\mu,g_0)=\delta_{ij} -\frac{g_0^2}{16 \pi^2} \Delta R_{ij}  (a\mu) ,
\label{eq:zlatcont}
\end{equation}
which renormalize the results of Monte Carlo simulations to a continuum scheme.

We choose as continuum scheme the $\overline{\rm{MS}}$ scheme, since commonly
the Wilson coefficients are computed in this scheme. On the lattice the 
renormalization condition is that the 1-loop amputated matrix elements 
at a certain reference scale $\mu$ are equal to the corresponding bare 
tree--level quantities. For lattice matrix elements of multiplicatively 
renormalized operators computed between one-quark states, this condition reads
\begin{eqnarray}
\langle p |O^{\rm lat}(\mu) | p \rangle \Big|_{p^2=\mu^2} & = & Z_O 
(a\mu, g_0) \cdot Z^{-1}_\psi (a\mu, g_0) \cdot \langle p |O^{(0)}(a) 
| p  \rangle \Big|^{\rm 1-loop}_{p^2=\mu^2} \nonumber \\
& = & \langle p |O^{(0)}(a)   | p  \rangle 
\Big|^{\rm tree}_{p^2=\mu^2},
\end{eqnarray}
where $Z_\psi$ is the wave-function renormalization, computed from the
quark self-energy.

\begin{figure}[btp]
  \begin{center}
    \epsfig{file=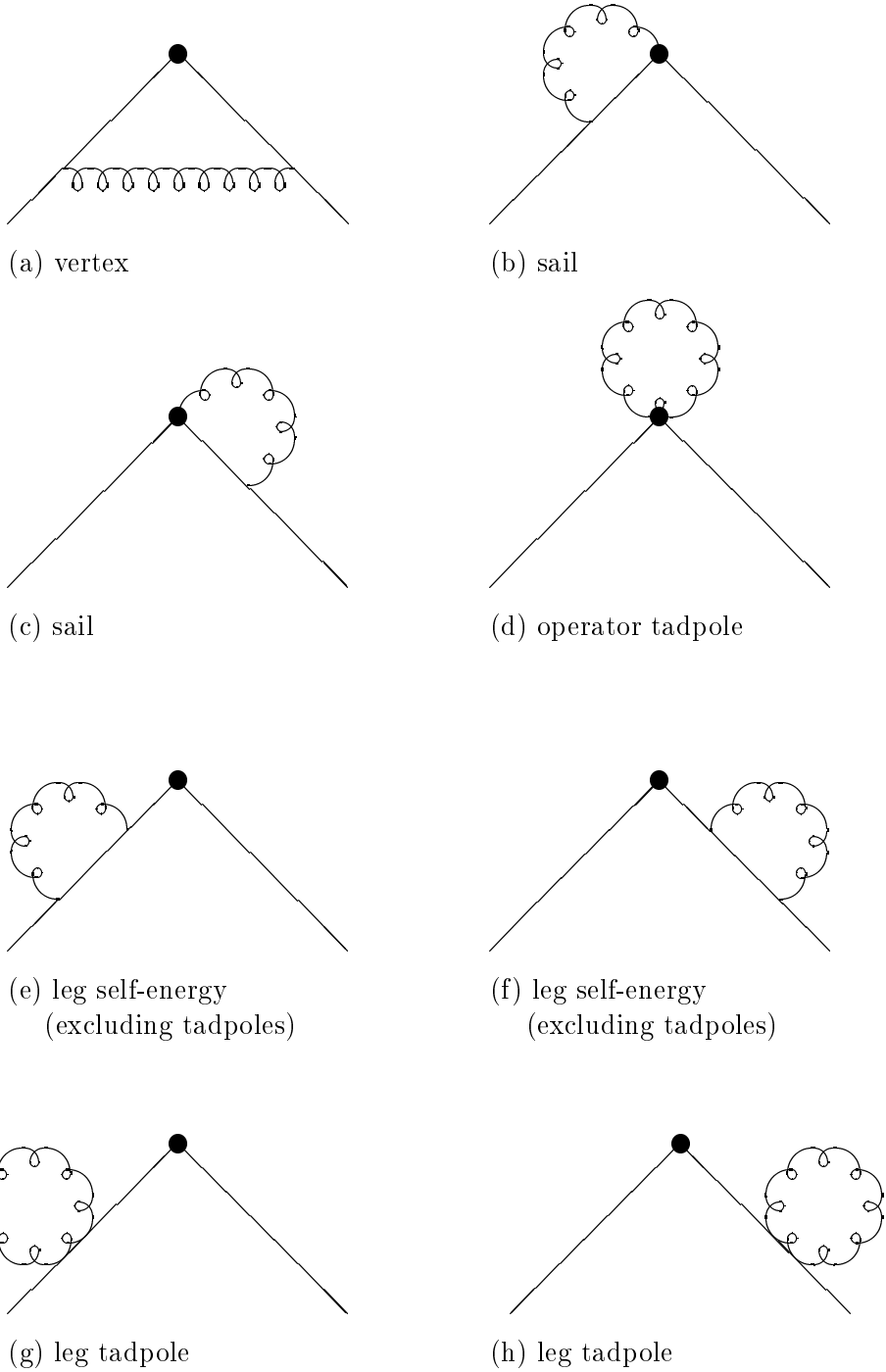, width=.99\linewidth}
    \caption{The graphs contributing to the 1-loop renormalization factors of 
       the matrix elements $\langle p| O |p \rangle$. The operator insertion 
       is indicated by a circle.}
\vspace{0.3cm}
    \label{fig:diagrams}
  \end{center}
\end{figure}

The 1-loop lattice matrix element of a logarithmically divergent operator 
$O$ has the form 
\begin{eqnarray}
\langle p |O^{(0)}(a) | p  \rangle \Big|^{\rm 1-loop} &=&
\langle p |O^{(0)}(a)   | p  \rangle \Big|^{\rm tree} \times \\
&& \qquad \Bigg( 1 + \frac{g_0^2}{16\pi^2} C_F \Big( \gamma_O \log a^2 p^2 
+V_O + T_O +2S \Big) \Bigg) \nonumber ,
\end{eqnarray}
where $V_O$ is the finite contribution of the vertex and sails diagrams 
(a, b and c in Fig.~1), $T_O$ refers to the tadpole arising from the 
operator (d in Fig.~1), and S is the finite contribution (proportional to 
${\rm i} p \!\!\! /$) of the quark self-energy of one leg, including the 
leg tadpole (e and g, 
or f and h, in Fig.~1). We have also that $C_F=\frac{N_c^2-1}{2N_c}$ for
the $SU(N_c)$ gauge group. 
The $Z_O$ factor for the operator $O$ is then given by
\begin{equation}
Z_O (a\mu, g_0) = 
1 - \frac{g_0^2}{16\pi^2} C_F \Bigg( \gamma_O \log a^2 \mu^2 +B_O \Bigg) ,
\label{eq:zeta1}
\end{equation}
with
\begin{equation}
B_O = V_O + T_O + S.
\label{eq:zeta2}
\end{equation}
We will call ``proper'' contributions the ones that do not include the 
self-energy diagrams. They correspond to the diagrams a-d in Fig.~1.

We have used computer programs to carry out the calculations of the Feynman 
diagrams, given their complexities and the huge number of terms that are 
present in some cases. They utilize an ensemble of routines written in 
the symbolic manipulation language FORM, and are able to do the analytic 
calculations; Fortran codes subsequently perform the numerical integrations. 
These routines are an extension of the ones used to do calculations with 
the Wilson action in various occasions~\cite{calcoli,calcoli2} and to compute 
the first two moments of the momentum and helicity distributions with overlap 
fermions~\cite{primidue}. Some more details can be found in 
Ref.~\cite{primidue}. 

We would like to stress that one of the major checks of our calculations, 
which are all done in a general covariant gauge~\footnote{The gluon propagator
that we use is 
\begin{equation}
G_{\mu\nu}(k) = \frac{1}{4\sum_\rho \sin^2 \frac{k_\rho}{2}}
\Bigg( \delta_{\mu\nu} - (1-\alpha) \frac{4 \sin \frac{k_\mu}{2} 
\sin \frac{k_\nu}{2}}{4\sum_\lambda \sin^2 \frac{k_\lambda}{2}} \Bigg) .
\end{equation}
}, is the cancellation of the gauge-dependent terms proportional to 
$(1-\alpha)$ in the final numbers that connect the lattice to the 
$\overline{\rm{MS}}$ scheme, as in Eq.~(\ref{eq:zlatcont}). 
Another significant check with overlap fermions is that the operators 
$O_{v_4,d}$ and $O_{a_4,d}$ have the same $Z$ well within numerical 
integration errors, 
as well as $O_{v_4,e}$ and $O_{a_4,e}$, as expected from chiral symmetry. 
Furthermore, we verified that there is no $1/a$ mixing term in the 1-loop 
expressions of the operators $O_{d_1}$, $O_{d_2}$ and $O_{d_3}$, contrary 
to what happens in the Wilson case.

The computations of the integrals for the operators $O_{v_4,e}$, $O_{a_4,e}$ 
and $O_{d_3}$ have been the most demanding. We had to split the FORM outputs 
in several pieces as otherwise the Fortran programs would refuse to compile 
them. The running time for the computation of the integrals on a $60^4$ grid 
has been in these cases of the order of 500 hours on a 500 MHz CPU.

\section{Improvement: overlap versus Wilson fermions}

To highlight the advantages of overlap fermions relative to Wilson fermions,
we summarize the differences in $O(a)$ improvement using each method.

Although the overlap-Dirac action possesses an exact chiral symmetry and has
no $O(a)$ terms, matrix elements of operators do have order $a$ corrections 
and therefore they need to be improved. The operators considered in this 
paper, which are all of the form $O = \bar{\psi} \widetilde{O} \psi$, are 
made free of $O(a)$ corrections by taking~\cite{cghrs}
\begin{equation}
O^{\rm imp} = \bar{\psi} \Big(1-\frac{1}{2\rho} a D_N\Big) \, \widetilde{O}
\, \Big(1-\frac{1}{2\rho} a D_N\Big) \psi.
\label{eq:opimp}
\end{equation}
Thus, in the overlap formulation, by rotating the spinors as in 
Eq.~(\ref{eq:opimp}), operators are freed of $O(a)$ corrections. In 
Ref.~\cite{calcoli2} it has been shown that this recipe improves the 
operators to all orders of perturbation theory. Thus, full $O(a)$ 
improvement is achieved without tuning any coefficients. Furthermore, 
the renormalization constants for the improved and unimproved operators 
are the same, so there is no need to do additional calculations, which 
are generally cumbersome, to compute the renormalization factors in the 
improved theory. What happens is that in 1-loop amplitudes a factor $D_N$ 
can combine with a quark propagator, but since it has an $a$ in front and 
(contrary to the Wilson case) there is no $1/a$ piece in the propagator, 
as additive mass renormalization is forbidden by chiral symmetry, the 
contribution of $D_N$ to the renormalization factors is zero~\cite{afpv}. 
Thus, improved Monte Carlo simulations are performed with the 
operator~(\ref{eq:opimp}), nevertheless the renormalization of this
operator is equal to that of the corresponding unimproved operator.

For Wilson fermions, instead, the improvement of operators looks more
complicated and troublesome. First of all, one has to find out a complete 
basis which includes all operators which are one-dimension higher than the 
original one and have the same symmetries. Then, to get the improved 
renormalization factors, which are different from the unimproved ones, 
one has to compute the 1-loop matrix elements of each one of those operators. 
Lastly, one has to determine somehow the values of the coefficients in front 
of the operator counterterms, and this appears to be a highly non-trivial task.

Let us consider, for example, the first moment of the quark momentum 
distribution, $O_{v_2} = \bar{\psi} \gamma_{\{\mu} D_{\nu\}} \psi$. 
The improvement counterterms are two dimension-five operators~\cite{calcoli2},
and one basis for the improvement is given by
\begin{equation}
 \bar{\psi} \gamma_{\{\mu} D_{\nu\}} \psi
-\frac{1}{4} a {\rm i} c_1 (g_0^2) \, \sum_\lambda 
   \bar{\psi}\sigma_{\lambda\{\mu} \Big[ D_{\nu\}} , D_\lambda \Big] \psi  
  -\frac{1}{4} a c_2 (g_0^2) \, \bar{\psi} \Big\{ D_\mu , D_\nu 
   \Big\} \psi .
\end{equation}
The operator is fully $O(a)$ improved only for some particular values 
of the coefficients $c_1 (g_0^2) = 1 + g_0^2 c_1^{(1)} + O(g_0^4) $ and 
$c_2 (g_0^2) = 1 + g_0^2 c_2^{(1)} + O(g_0^4) $. However, presently one knows 
only a relation between the two coefficients~\cite{calcoli2}, and thus one 
of them remains unknown. Although one could determine all coefficients using 
some Ward Identities or physical conditions, this involves more effort than 
with overlap fermions. In addition, the coefficients are different for 
different orders in perturbation theory, and the basis itself is different 
for each operator to be improved.
It is reasonable to expect that operators which contain more covariant 
derivatives will have larger improvement bases, and thus many more 
coefficients to be determined, besides computing the 1-loop matrix elements
for each of the operator counterterms. Presumably already implementing the 
improvement for the second moment operators with Wilson fermions will turn 
out to be a daunting task. This means that in most cases it will be difficult 
to go beyond tree-level improvement for the structure function operators 
in the Wilson case.

In addition to improving operators, one should also note that Wilson fermions 
need corrections to their action (the Sheikholeslami-Wohlert term), while 
the overlap action is already $O(a)$ improved. However, this is less of a 
complication than the improvement of the operators. 

Thus, for all these reasons, from the perspective of renormalization, mixing,
and improvement, overlap fermions offer compelling advantages relative to
Wilson fermions.

\section{Results}

\begin{table}[hbtp]
\begin{center}
\vspace{0.5cm}
  \begin{tabular}[btp]{|c||r|r|r|r|r|r|r|r|} \hline
$\rho$ & $V_{v_4,d}^{\ \alpha=1}$ & $V_{v_4,e}^{\ \alpha=1}$ 
& $V_{d_1}^{\ \alpha=1}$ & $V_{d_2}^{\ \alpha=1}$ & $V_{d_3}^{\ \alpha=1}$ 
& $V_{t_1}^{\ \alpha=1}$ & $V_{t_2}^{\ \alpha=1}$ & $V_{t_3}^{\ \alpha=1}$ 
\\ \hline
0.2  &  -9.61600  &  -7.98774  &  52.66717  &  45.13633  &  41.92810  
     &   5.38049  &  -4.02560  &  -7.79945  \\
0.3  &  -9.46984  &  -7.84341  &  33.37857  &  26.64726  &  23.65026  
     &   5.19029  &  -4.00839  &  -7.70244  \\
0.4  &  -9.34521  &  -7.72067  &  24.03566  &  17.76727  &  14.91522  
     &   5.03112  &  -3.99077  &  -7.61751  \\
0.5  &  -9.23484  &  -7.61223  &  18.60963  &  12.65715  &   9.91543  
     &   4.89210  &  -3.97274  &  -7.54054  \\
0.6  &  -9.13474  &  -7.51412  &  15.11256  &   9.39617  &   6.74341  
     &   4.76733  &  -3.95427  &  -7.46925  \\
0.7  &  -9.04244  &  -7.42384  &  12.70136  &   7.17180  &   4.59344  
     &   4.65325  &  -3.93535  &  -7.40221  \\
0.8  &  -8.95626  &  -7.33973  &  10.95882  &   5.58297  &   3.06844  
     &   4.54744  &  -3.91595  &  -7.33847  \\
0.9  &  -8.87501  &  -7.26060  &   9.65553  &   4.40974  &   1.95101  
     &   4.44823  &  -3.89606  &  -7.27733  \\
1.0  &  -8.79780  &  -7.18557  &   8.65526  &   3.52185  &   1.11258  
     &   4.35437  &  -3.87564  &  -7.21828  \\
1.1  &  -8.72396  &  -7.11394  &   7.87218  &   2.83746  &   0.47256  
     &   4.26491  &  -3.85467  &  -7.16091  \\
1.2  &  -8.65295  &  -7.04517  &   7.24967  &   2.30276  &  -0.02200  
     &   4.17912  &  -3.83313  &  -7.10491  \\
1.3  &  -8.58433  &  -6.97884  &   6.74895  &   1.88097  &  -0.40718  
     &   4.09639  &  -3.81099  &  -7.05001  \\
1.4  &  -8.51774  &  -6.91457  &   6.34259  &   1.54618  &  -0.70839  
     &   4.01623  &  -3.78821  &  -6.99600  \\
1.5  &  -8.45289  &  -6.85207  &   6.01069  &   1.27965  &  -0.94394  
     &   3.93824  &  -3.76477  &  -6.94269  \\
1.6  &  -8.38952  &  -6.79107  &   5.73851  &   1.06752  &  -1.12735  
     &   3.86206  &  -3.74064  &  -6.88993  \\
1.7  &  -8.32742  &  -6.73135  &   5.51486  &   0.89934  &  -1.26882  
     &   3.78741  &  -3.71578  &  -6.83758  \\
1.8  &  -8.26640  &  -6.67272  &   5.33115  &   0.76710  &  -1.37612  
     &   3.71403  &  -3.69017  &  -6.78553  \\
\hline
  \end{tabular} 
\vspace{0.5cm}
  \caption{The Feynman-gauge constants $V^{\ \alpha=1}_O$ for the
momentum, helicity and transversity operators considered in this work,
in the overlap theory. Note that in this case 
$V_{a_4,d}^{\ \alpha=1}=V_{v_4,d}^{\ \alpha=1}$ and
$V_{a_4,e}^{\ \alpha=1}=V_{v_4,e}^{\ \alpha=1}$.}
\end{center}
\vspace{0.5cm}
\end{table}

\begin{table}[hbtp]
\begin{center}
\vspace{0.5cm}
  \begin{tabular}[btp]{|c||l|} \hline
 operator & operator tadpole \\ \hline
& \\
$O_{v_4,d},O_{a_4,d}$ & $ T_{v_4,d} = T_{a_4,d} = 
16 \pi^2 \Big( -\displaystyle{\frac{3}{2}} Z_0 + (1-\alpha) \Big(
-\displaystyle{\frac{1}{24}} Z_0 -\displaystyle{\frac{1}{4}} Z_1 
+\displaystyle{\frac{1}{16}} \Big) \Big) $ \\
& \\
$O_{v_4,e},O_{a_4,e}$ & $ T_{v_4,e} = T_{a_4,e} = 
16 \pi^2 \Big( -\displaystyle{\frac{5}{2}} Z_0 -\displaystyle{\frac{1}{3}} Z_1
+\displaystyle{\frac{1}{6}}
+ (1-\alpha) \Big(\displaystyle{\frac{9}{8}} Z_0 +Z_1 
-\displaystyle{\frac{1}{4}} \Big) \Big) $ \\
& \\
$O_{d_1}$ & $ T_{d_1} = 16 \pi^2 \Big( -\displaystyle{\frac{1}{2}} Z_0 
+ (1-\alpha) \displaystyle{\frac{1}{8}} Z_0 \Big) $ \\
& \\
$O_{d_2}$ & $ T_{d_2} = 16 \pi^2 \Big( -Z_0 + (1-\alpha) 
\displaystyle{\frac{1}{6}} Z_0 \Big) $ \\
& \\
$O_{d_3}$ & $ T_{d_3} = 16 \pi^2 \Big( -\displaystyle{\frac{3}{2}} Z_0 
+ (1-\alpha) \Big( -\displaystyle{\frac{1}{24}} Z_0 
-\displaystyle{\frac{1}{4}} Z_1 +\displaystyle{\frac{1}{16}} \Big) \Big) $ \\
& \\
$O_{t_1}$ & $ T_{t_1} = 0$ \\
& \\
$O_{t_2}$ & $ T_{t_2} = 16 \pi^2 \Big( -\displaystyle{\frac{1}{2}} Z_0 
+ (1-\alpha) \displaystyle{\frac{1}{8}} Z_0 \Big) $ \\
& \\
$O_{t_3}$ & $ T_{t_3} = 16 \pi^2 \Big( -Z_0 + (1-\alpha) 
\displaystyle{\frac{1}{6}} Z_0 \Big) $ \\
& \\
\hline
  \end{tabular} 
\vspace{0.5cm}
  \caption{The operator tadpoles for the various operators, where 
$Z_0=0.154933390231\cdots $ and $Z_1=0.107781313540\cdots $.}
\end{center}
\vspace{0.5cm}
\end{table}

We give in this section the results for the renormalization factors of every 
operator considered in this paper, both for the overlap and the Wilson action, 
for $r=1$. 

In the overlap case, we have explicitly verified that the renormalization
constants of $O_{a_4,d}$ and $O_{a_4,e}$ are equal to the ones of $O_{v_4,d}$
and $O_{v_4,e}$ respectively, which are given below. This is a consequence
of the exact chiral symmetry which the overlap theory possesses. 
Another important gain coming from chiral symmetry is that the operators 
$O_{d_1}$, $O_{d_2}$ and $O_{d_3}$ are multiplicatively renormalized, 
while for Wilson fermions the breaking of chiral symmetry causes a mixing 
with operators of lower dimensions, with the mixing coefficients going to
infinity in the continuum limit.

\subsection{Overlap fermions}

We first consider the 1-loop contributions of the proper diagrams (a-d in 
Fig.~1), which are
\begin{eqnarray}
O_{v_4,d}^{\rm proper} =\frac{g_0^2}{16\pi^2} C_F && \Bigg[ \Big( 
\frac{127}{30} + (1-\alpha) \Big) \log a^2 p^2  \nonumber \\ 
&& \qquad + V^{\ \alpha=1}_{v_4,d} - (1-\alpha)\, 7.553824 
+ T_{v_4,d} \Bigg] O_{v_4,d}^{\rm tree} , \nonumber \\
O_{v_4,e}^{\rm proper} =\frac{g_0^2}{16\pi^2} C_F && \Bigg[ \Big( 
\frac{127}{30} + (1-\alpha) \Big) \log a^2 p^2 \\ 
&& \qquad + V^{\ \alpha=1}_{v_4,e} - (1-\alpha)\, 8.024764
+ T_{v_4,e} \Bigg] O_{v_4,e}^{\rm tree} , \nonumber \\
O_{d_1}^{\rm proper} =\frac{g_0^2}{16\pi^2} C_F && \Bigg[ -\alpha \log a^2 p^2 
+ V^{\ \alpha=1}_{d_1} - (1-\alpha)\, 7.850272
+ T_{d_1} \Bigg] O_{d_1}^{\rm tree} , \nonumber \\
O_{d_2}^{\rm proper} =\frac{g_0^2}{16\pi^2} C_F && \Bigg[ \Big( 
\frac{1}{6} + (1-\alpha) \Big) \log a^2 p^2 \nonumber \\ 
&& \qquad + V^{\ \alpha=1}_{d_2} - (1-\alpha)\, 8.369693
+ T_{d_2} \Bigg] O_{d_2}^{\rm tree} \nonumber \\
O_{d_3}^{\rm proper} =\frac{g_0^2}{16\pi^2} C_F && \Bigg[ \Big( 
\frac{17}{18} + (1-\alpha) \Big) \log a^2 p^2 \nonumber \\ 
&& \qquad + V^{\ \alpha=1}_{d_3} - (1-\alpha)\, 8.553824
+ T_{d_3} \Bigg] O_{d_3}^{\rm tree} \nonumber \\
O_{t_1}^{\rm proper} =\frac{g_0^2}{16\pi^2} C_F && \Bigg[
(1-\alpha) \log a^2 p^2 \nonumber \\ 
&& \qquad + V^{\ \alpha=1}_{t_1} - (1-\alpha)\, 3.792010
+ T_{t_1} \Bigg] O_{t_1}^{\rm tree} \nonumber \\
O_{t_2}^{\rm proper} =\frac{g_0^2}{16\pi^2} C_F && \Bigg[ \Big( 
2 + (1-\alpha) \Big) \log a^2 p^2 \nonumber \\ 
&& \qquad + V^{\ \alpha=1}_{t_2} - (1-\alpha)\, 6.350272
+ T_{t_2} \Bigg] O_{t_2}^{\rm tree} \nonumber \\
O_{t_3}^{\rm proper} =\frac{g_0^2}{16\pi^2} C_F && \Bigg[ \Big( 
\frac{10}{3} + (1-\alpha) \Big) \log a^2 p^2 \nonumber \\ 
&& \qquad + V^{\ \alpha=1}_{t_3} - (1-\alpha)\, 7.036360
+ T_{t_3} \Bigg] O_{t_3}^{\rm tree} \nonumber .
\end{eqnarray}
The Feynman gauge results $V^{\ \alpha=1}_O$ for the contributions of the 
sails and vertices, $V_O$, are tabulated in Table 1 for several values 
of the parameter $\rho$. They are also given separately for each diagram 
in Appendix B (for $\rho=1$). 
The remaining parts proportional to $(1-\alpha)$ are instead independent of 
$\rho$, and as their analytic expressions are very complicated functions of 
$\rho$ containing thousands of terms, the numerical cancellation of this 
dependence is a highly non-trivial check of our computations. Furthermore, the 
parts proportional to $(1-\alpha)$ have also the same value as with the Wilson
action~\cite{primidue}. In fact they depend only on the gluonic action chosen.

The result for the tensor charge 
$O_{t_1} = \bar{\psi} \sigma_{\mu\nu} \gamma_5 \psi $ turns out to be equal 
to the result for the standard tensor current 
$\bar{\psi} \sigma_{\mu\nu} \psi $, which has been calculated in 
Ref.~\cite{afpv} in Feynman gauge and in Ref.~\cite{cg} in a general 
covariant gauge. For Wilson fermions, the renormalization of $O_{t_1}$ 
can be found in~\cite{calcoli2}.

The results for the remaining proper diagram, the operator tadpole $T_O$ 
(diagram d in Fig.~1), are shown in Table 2. To compute the renormalization 
factors, one finally adds to the proper diagrams the 1-loop amplitudes of 
the self-energy and tadpole of one leg which are proportional to 
${\rm i} p \!\!\! /$. Their values are given in Appendix A.
Putting everything together, we get the expressions of the renormalized 
operators on the lattice for overlap fermions, which for $\rho=1$ are:
\begin{eqnarray}
\widehat{O}_{v_4,d} 
 &=&\Bigg[ 1 -\frac{g_0^2}{16\pi^2} C_F \Bigg(\frac{157}{30} \log a^2 \mu^2 
-83.12758 +\frac{11}{6}
(1-\alpha) \Bigg) \Bigg] O_{v_4,d}^{\rm tree} \nonumber \\ 
\widehat{O}_{v_4,e} 
&=&\Bigg[ 1 -\frac{g_0^2}{16\pi^2} C_F \Bigg(\frac{157}{30} \log a^2 \mu^2 
-85.33588 +\frac{11}{6}
(1-\alpha) \Bigg) \Bigg] O_{v_4,e}^{\rm tree} \nonumber \\
\widehat{O}_{d_1} 
 &=&\Bigg[ 1 +\frac{g_0^2}{16\pi^2} C_F \cdot 41.20842 \Bigg] 
O_{d_1}^{\rm tree} \label{eq:ovres}\\ 
\widehat{O}_{d_2} 
&=&\Bigg[ 1 -\frac{g_0^2}{16\pi^2} C_F \Bigg(\frac{7}{6} \log a^2 \mu^2 
-58.57488 +\frac{1}{2}(1-\alpha) \Bigg) \Bigg] O_{d_2}^{\rm tree} 
\nonumber \\
\widehat{O}_{d_3} 
&=&\Bigg[ 1 -\frac{g_0^2}{16\pi^2} C_F \Bigg(\frac{35}{18} \log a^2 \mu^2 
-73.21719 +\frac{5}{6}(1-\alpha) \Bigg) \Bigg] O_{d_3}^{\rm tree} 
\nonumber \\
\widehat{O}_{t_1} 
&=&\Bigg[ 1 -\frac{g_0^2}{16\pi^2} C_F \Bigg(\log a^2 \mu^2 
-33.27626 + (1-\alpha) \Bigg) \Bigg] O_{t_1}^{\rm tree} 
\nonumber \\
\widehat{O}_{t_2} 
&=&\Bigg[ 1 -\frac{g_0^2}{16\pi^2} C_F \Bigg(3\log a^2 \mu^2 
-53.73932 +\frac{3}{2}(1-\alpha) \Bigg) \Bigg] O_{t_2}^{\rm tree} 
\nonumber \\
\widehat{O}_{t_3} 
&=&\Bigg[ 1 -\frac{g_0^2}{16\pi^2} C_F \Bigg(\frac{13}{3} \log a^2 \mu^2 
-69.31500 +\frac{11}{6}(1-\alpha) \Bigg) \Bigg] O_{t_3}^{\rm tree} 
.\nonumber 
\end{eqnarray}
We see that in the overlap case the operators $O_{d_1}$, $O_{d_2}$ and 
$O_{d_3}$ are multiplicatively renormalized (contrary to what happens with 
Wilson fermions, see below).

Another check of our calculations is that the transversity operators, which 
apart from the trivial tensor charge have never been computed before on the 
lattice, agree with the 1-loop anomalous dimension formula~\cite{am}
\begin{equation}
\gamma_{t_n} = 1 + 4 \sum_{j=2}^n \frac{1}{j} , 
\end{equation}
which also implies that the anomalous dimensions are positive and for any 
given moment greater than the ones of the corresponding moments of 
$F_1$, $F_2$ and $g_1$.

To complete the connection with the continuum physics as in 
Eq.~(\ref{eq:matching}), we need also the 1-loop results for the same matrix 
elements in the continuum $\overline{\rm{MS}}$ scheme, which are given by 
\begin{eqnarray}
\widehat{O}^{\overline{\rm{MS}}}_{v_4}  
 &=&\Bigg[ 1 -\frac{g_0^2}{16\pi^2} C_F \Bigg(\frac{157}{30} \log a^2 \mu^2 
-\frac{2216}{225} +\frac{11}{6}
(1-\alpha) \Bigg) \Bigg] O_{v_4}^{\rm tree} \nonumber \\ 
\widehat{O}^{\overline{\rm{MS}}}_{a_4}  
 &=&\Bigg[ 1 -\frac{g_0^2}{16\pi^2} C_F \Bigg(\frac{157}{30} \log a^2 \mu^2 
-\frac{2216}{225} +\frac{11}{6}
(1-\alpha) \Bigg) \Bigg] O_{a_4}^{\rm tree} \nonumber \\ 
\widehat{O}^{\overline{\rm{MS}}}_{d_1} 
&=& O_{d_1}^{\rm tree} \\ 
\widehat{O}^{\overline{\rm{MS}}}_{d_2} 
&=&\Bigg[ 1 -\frac{g_0^2}{16\pi^2} C_F \Bigg(\frac{7}{6} \log a^2 \mu^2 
-\frac{35}{18} +\frac{1}{2}(1-\alpha) \Bigg) \Bigg] O_{d_2}^{\rm tree} 
\nonumber \\ 
\widehat{O}^{\overline{\rm{MS}}}_{d_3} 
&=&\Bigg[ 1 -\frac{g_0^2}{16\pi^2} C_F \Bigg(\frac{35}{18} \log a^2 \mu^2 
-\frac{92}{27} +\frac{5}{6}(1-\alpha) \Bigg) \Bigg] O_{d_3}^{\rm tree} 
\nonumber \\
\widehat{O}^{\overline{\rm{MS}}}_{t_1} 
&=&\Bigg[ 1 -\frac{g_0^2}{16\pi^2} C_F \Bigg(\log a^2 \mu^2 
-1 +(1-\alpha) \Bigg) \Bigg] O_{t_1}^{\rm tree} 
\nonumber \\ 
\widehat{O}^{\overline{\rm{MS}}}_{t_2} 
&=&\Bigg[ 1 -\frac{g_0^2}{16\pi^2} C_F \Bigg(3\log a^2 \mu^2 
-5 +\frac{3}{2}(1-\alpha) \Bigg) \Bigg] O_{t_2}^{\rm tree} 
\nonumber \\ 
\widehat{O}^{\overline{\rm{MS}}}_{t_3} 
&=&\Bigg[ 1 -\frac{g_0^2}{16\pi^2} C_F \Bigg(\frac{13}{3} \log a^2 \mu^2 
-\frac{71}{9} +\frac{11}{6}(1-\alpha) \Bigg) \Bigg] O_{t_3}^{\rm tree} 
. \nonumber 
\end{eqnarray}

The connection of overlap lattice fermions with the continuum 
$\overline{\rm{MS}}$ is then given by the gauge-invariant factors
\begin{eqnarray}
\widehat{O}^{\overline{\rm{MS}}}_{v_4,d} 
&=&\Bigg[ 1 -\frac{g_0^2}{16\pi^2} C_F \Bigg(\frac{157}{30} \log a^2 \mu^2 
-73.27869 \Bigg) \Bigg] O^{\rm lat~(overlap)}_{v_4,d} \nonumber \\ 
\widehat{O}^{\overline{\rm{MS}}}_{a_4,d} 
&=&\Bigg[ 1 -\frac{g_0^2}{16\pi^2} C_F \Bigg(\frac{157}{30} \log a^2 \mu^2 
-73.27869 \Bigg) \Bigg] O^{\rm lat~(overlap)}_{a_4,d} \nonumber \\ 
\widehat{O}^{\overline{\rm{MS}}}_{v_4,e} 
&=&\Bigg[ 1 -\frac{g_0^2}{16\pi^2} C_F \Bigg(\frac{157}{30} \log a^2 \mu^2 
-75.48699 \Bigg) \Bigg] O^{\rm lat~(overlap)}_{v_4,e} \nonumber \\ 
\widehat{O}^{\overline{\rm{MS}}}_{a_4,e} 
&=&\Bigg[ 1 -\frac{g_0^2}{16\pi^2} C_F \Bigg(\frac{157}{30} \log a^2 \mu^2 
-75.48699 \Bigg) \Bigg] O^{\rm lat~(overlap)}_{a_4,e} \nonumber \\ 
\widehat{O}^{\overline{\rm{MS}}}_{d_1} 
&=&\Bigg[ 1 +\frac{g_0^2}{16\pi^2} C_F \cdot 41.20842 \Bigg) \Bigg] 
O^{\rm lat~(overlap)}_{d_1} \label{eq:overlaptomsbar}\\ 
\widehat{O}^{\overline{\rm{MS}}}_{d_2} 
&=&\Bigg[ 1 -\frac{g_0^2}{16\pi^2} C_F \Bigg(\frac{7}{6} \log a^2 \mu^2 
-56.63044 \Bigg) \Bigg] O^{\rm lat~(overlap)}_{d_2} \nonumber \\ 
\widehat{O}^{\overline{\rm{MS}}}_{d_3} 
&=&\Bigg[ 1 -\frac{g_0^2}{16\pi^2} C_F \Bigg(\frac{35}{18} \log a^2 \mu^2 
-69.80979 \Bigg) \Bigg] O^{\rm lat~(overlap)}_{d_3} \nonumber \\ 
\widehat{O}^{\overline{\rm{MS}}}_{t_1} 
&=&\Bigg[ 1 -\frac{g_0^2}{16\pi^2} C_F \Bigg(\log a^2 \mu^2 
-32.27626 \Bigg) \Bigg] O^{\rm lat~(overlap)}_{t_1} \nonumber \\ 
\widehat{O}^{\overline{\rm{MS}}}_{t_2} 
&=&\Bigg[ 1 -\frac{g_0^2}{16\pi^2} C_F \Bigg(3 \log a^2 \mu^2 
-48.73932 \Bigg) \Bigg] O^{\rm lat~(overlap)}_{t_2} \nonumber \\ 
\widehat{O}^{\overline{\rm{MS}}}_{t_3} 
&=&\Bigg[ 1 -\frac{g_0^2}{16\pi^2} C_F \Bigg(\frac{13}{3} \log a^2 \mu^2 
-61.42612 \Bigg) \Bigg] O^{\rm lat~(overlap)}_{t_3} . \nonumber 
\end{eqnarray}
For $\beta=6.0$, $\mu=1/a$ and $N_c=3$ one has
\begin{eqnarray}
\widehat{O}^{\overline{\rm{MS}}}_{v_4,d} 
&=& 1.61872 \, O^{\rm lat~(overlap)}_{v_4,d}
  = 1.19722 \, O^{\rm lat~(Wilson)}_{v_4,d} \nonumber \\ 
\widehat{O}^{\overline{\rm{MS}}}_{a_4,d} 
&=& 1.61872 \, O^{\rm lat~(overlap)}_{a_4,d}  
  = 1.20040 \, O^{\rm lat~(Wilson)}_{a_4,d} \nonumber \\ 
\widehat{O}^{\overline{\rm{MS}}}_{v_4,e} 
&=& 1.63737 \, O^{\rm lat~(overlap)}_{v_4,e} 
  = 1.21534 \, O^{\rm lat~(Wilson)}_{v_4,e} \nonumber \\ 
\widehat{O}^{\overline{\rm{MS}}}_{a_4,e} 
&=& 1.63737 \, O^{\rm lat~(overlap)}_{a_4,e} 
  = 1.21944 \, O^{\rm lat~(Wilson)}_{a_4,e} \nonumber \\ 
\widehat{O}^{\overline{\rm{MS}}}_{d_1} 
&=& 1.34794 \, O^{\rm lat~(overlap)}_{d_1} \label{eq:overlapwilsonmsbar}\\ 
\widehat{O}^{\overline{\rm{MS}}}_{d_2} 
&=& 1.47816 \, O^{\rm lat~(overlap)}_{d_2} \nonumber \\ 
\widehat{O}^{\overline{\rm{MS}}}_{d_3} 
&=& 1.58943 \, O^{\rm lat~(overlap)}_{d_3} \nonumber \\ 
\widehat{O}^{\overline{\rm{MS}}}_{t_1} 
&=& 1.27252 \, O^{\rm lat~(overlap)}_{t_1}  
  = 0.85631 \, O^{\rm lat~(Wilson)}_{t_1} \nonumber \\ 
\widehat{O}^{\overline{\rm{MS}}}_{t_2} 
&=& 1.41153 \, O^{\rm lat~(overlap)}_{t_2} 
  = 0.99559 \, O^{\rm lat~(Wilson)}_{t_2} \nonumber \\ 
\widehat{O}^{\overline{\rm{MS}}}_{t_3} 
&=& 1.51865 \, O^{\rm lat~(overlap)}_{t_3}  
  = 1.10021 \, O^{\rm lat~(Wilson)}_{t_3} , \nonumber 
\end{eqnarray}
where we have also shown the corresponding Wilson results (see 
Eq.~(\ref{eq:wilsontomsbar})). We remind that, although when $\mu=1/a$ the 
logarithms disappear from the renormalization factors, for a general $\mu$ 
they are still present but they cancel then against the corresponding 
logarithms in the Wilson coefficients, as the moments Eq.~(\ref{eq:momsf}) 
have to be independent of the renormalization scale (and all the scale 
dependency is in the logarithms).

The renormalization constants for overlap fermions reported here look in 
general large, especially when compared to the corresponding Wilson results, 
as shown in Eq.~(\ref{eq:overlapwilsonmsbar}). However, if one looks at the 
overlap results for the proper diagrams (see for example Appendix B), one can 
notice that they are not so much different from Wilson fermions. In general, 
the biggest contribution to the renormalization constants comes from the 
operator tadpoles, but as they are exactly the same for overlap and Wilson 
fermions, it is not in these diagrams that the difference can be found. 
One instead has to look at the quark self-energy (see Appendix A). 
In the Feynman gauge, the leg self-energy in the overlap (for $\rho=1$) 
is --37.63063, while for Wilson fermions is +11.85240; their difference 
is --49.48303, and it is remarkable how close this number comes to the 
differences between the complete finite contributions $B_O$ (defined in 
Eqs.~(\ref{eq:zeta1}) and (\ref{eq:zeta2})) for the two kinds of fermions,
Eq.~(\ref{eq:overlaptomsbar}) vs. Eq.~(\ref{eq:wilsontomsbar}). 
If one would consider the overlap for $\rho=1.9$, the difference between
the self-energies would go from --49.48303 down to --26.80898; however, the 
quark propagator becomes singular for $\rho=2$, so simulations would likely 
be more expensive when approaching this value of $\rho$.

\subsection{Wilson fermions}

The 1-loop contributions of the proper diagrams are in the Wilson case
\begin{eqnarray}
O_{v_4,d}^{\rm proper} =\frac{g_0^2}{16\pi^2} C_F && \Bigg[ \Big( 
\frac{127}{30} + (1-\alpha) \Big) \log a^2 p^2  \nonumber \\ 
&& \qquad  -8.359667 - (1-\alpha)\, 7.553824 
+ T_{v_4,d} \Bigg] O_{v_4,d}^{\rm tree} , \nonumber \\
O_{a_4,d}^{\rm proper} =\frac{g_0^2}{16\pi^2} C_F && \Bigg[ \Big( 
\frac{127}{30} + (1-\alpha) \Big) \log a^2 p^2  \nonumber \\ 
&& \qquad  -8.736011 - (1-\alpha)\, 7.553824 
+ T_{a_4,d} \Bigg] O_{a_4,d}^{\rm tree} , \nonumber \\
O_{v_4,e}^{\rm proper} =\frac{g_0^2}{16\pi^2} C_F && \Bigg[ \Big( 
\frac{127}{30} + (1-\alpha) \Big) \log a^2 p^2 \nonumber \\ 
&& \qquad  -6.684639 - (1-\alpha)\, 8.024764
+ T_{v_4,e} \Bigg] O_{v_4,e}^{\rm tree} , \nonumber \\
O_{a_4,e}^{\rm proper} =\frac{g_0^2}{16\pi^2} C_F && \Bigg[ \Big( 
\frac{127}{30} + (1-\alpha) \Big) \log a^2 p^2 \\ 
&& \qquad  -7.171210 - (1-\alpha)\, 8.024764
+ T_{a_4,e} \Bigg] O_{a_4,e}^{\rm tree} , \nonumber \\
O_{d_1}^{\rm proper} =\frac{g_0^2}{16\pi^2} C_F && \Bigg[ -\alpha \log a^2 p^2 
+0.745643 - (1-\alpha)\, 7.850272
+ T_{d_1} \Bigg] O_{d_1}^{\rm tree} \nonumber \\
+16.243762 && \frac{\rm i}{a} \, \frac{g_0^2}{16\pi^2} C_F \,
\bar{\psi} \sigma_{41} \gamma_5 \psi  
\nonumber \\
O_{d_2}^{\rm proper} =\frac{g_0^2}{16\pi^2} C_F && \Bigg[ \Big( 
\frac{1}{6} + (1-\alpha) \Big) \log a^2 p^2 \nonumber \\ 
&& \qquad  -3.063751 - (1-\alpha)\, 8.369693
+ T_{d_2} \Bigg] O_{d_2}^{\rm tree} \nonumber \\
+4.265680 && \frac{1}{a} \, \frac{g_0^2}{16\pi^2} C_F \Big( 
  \bar{\psi} \gamma_4 \gamma_5 \gamma_{\{1} D_{2\}} \psi 
 -\bar{\psi} \gamma_1 \gamma_5 \gamma_{\{4} D_{2\}} \psi \Big) \nonumber \\
O_{d_3}^{\rm proper} =\frac{g_0^2}{16\pi^2} C_F && \Bigg[ \Big( 
\frac{17}{18} + (1-\alpha) \Big) \log a^2 p^2 \nonumber \\ 
&& \qquad  -4.902360 - (1-\alpha)\, 8.553824
+ T_{d_3} \Bigg] O_{d_3}^{\rm tree} \nonumber \\
+2.881820 && \frac{1}{a} \, \frac{g_0^2}{16\pi^2} C_F \Big( 
  \bar{\psi} \gamma_4 \gamma_5 \gamma_{\{1} D_2 D_{3\}} \psi 
 -\bar{\psi} \gamma_1 \gamma_5 \gamma_{\{4} D_2 D_{3\}} \psi \Big) \nonumber \\
O_{t_1}^{\rm proper} =\frac{g_0^2}{16\pi^2} C_F && \Bigg[
(1-\alpha) \log a^2 p^2 \nonumber \\ 
&& \qquad  +4.165675 - (1-\alpha)\, 3.792010
+ T_{t_1} \Bigg] O_{t_1}^{\rm tree} \nonumber \\
O_{t_2}^{\rm proper} =\frac{g_0^2}{16\pi^2} C_F && \Bigg[ \Big( 
2 + (1-\alpha) \Big) \log a^2 p^2 \nonumber \\ 
&& \qquad  -4.096894 - (1-\alpha)\, 6.350272
+ T_{t_2} \Bigg] O_{t_2}^{\rm tree} \nonumber \\
O_{t_3}^{\rm proper} =\frac{g_0^2}{16\pi^2} C_F && \Bigg[ \Big( 
\frac{10}{3} + (1-\alpha) \Big) \log a^2 p^2 \nonumber \\ 
&& \qquad  -7.143946 - (1-\alpha)\, 7.036360
+ T_{t_3} \Bigg] O_{t_3}^{\rm tree} , \nonumber 
\end{eqnarray}
where $\sigma_{\mu\nu}=\frac{\rm i}{2}[\gamma_\mu,\gamma_\nu]$.

The operator tadpoles are the same as in the overlap case (see Table 2). 
Adding the quark self-energy (appendix A), the complete renormalization 
factors are then
\begin{eqnarray}
\widehat{O}^{\rm Wilson}_{v_4,d} 
&=&\Bigg[ 1 -\frac{g_0^2}{16\pi^2} C_F \Bigg(\frac{157}{30} \log a^2 \mu^2 
-33.206413 +\frac{11}{6}
(1-\alpha) \Bigg) \Bigg] O_{v_4,d}^{\rm tree} \nonumber \\ 
\widehat{O}^{\rm Wilson}_{a_4,d} 
&=&\Bigg[ 1 -\frac{g_0^2}{16\pi^2} C_F \Bigg(\frac{157}{30} \log a^2 \mu^2 
-33.582757 +\frac{11}{6}
(1-\alpha) \Bigg) \Bigg] O_{a_4,d}^{\rm tree} \nonumber \\ 
\widehat{O}^{\rm Wilson}_{v_4,e} 
&=&\Bigg[ 1 -\frac{g_0^2}{16\pi^2} C_F \Bigg(\frac{157}{30} \log a^2 \mu^2 
-35.351922 +\frac{11}{6}
(1-\alpha) \Bigg) \Bigg] O_{v_4,e}^{\rm tree} \nonumber \\
\widehat{O}^{\rm Wilson}_{a_4,e} 
&=&\Bigg[ 1 -\frac{g_0^2}{16\pi^2} C_F \Bigg(\frac{157}{30} \log a^2 \mu^2 
-35.838493 +\frac{11}{6}
(1-\alpha) \Bigg) \Bigg] O_{a_4,e}^{\rm tree} \nonumber \\
\widehat{O}^{\rm Wilson}_{d_1} 
&=&\Bigg[ 1 -\frac{g_0^2}{16\pi^2} C_F \cdot 0.364997 \Bigg] 
O_{d_1}^{\rm tree} \nonumber \\
&& \qquad \qquad 
-16.243762 \, \frac{\rm i}{a} \, \frac{g_0^2}{16\pi^2} C_F \,
\bar{\psi} \sigma_{41} \gamma_5 \psi \nonumber \\
\widehat{O}^{\rm Wilson}_{d_2} 
&=&\Bigg[ 1 -\frac{g_0^2}{16\pi^2} C_F \Bigg(\frac{7}{6} \log a^2 \mu^2 
-15.677447  +\frac{1}{2}(1-\alpha) \Bigg) \Bigg] O_{d_2}^{\rm tree} 
\\
&& \qquad \qquad -4.265680 \, \frac{1}{a} \, \frac{g_0^2}{16\pi^2} C_F \Big( 
  \bar{\psi} \gamma_4 \gamma_5 \gamma_{\{1} D_{2\}} \psi 
 -\bar{\psi} \gamma_1 \gamma_5 \gamma_{\{4} D_{2\}} \psi \Big) \nonumber \\
\widehat{O}^{\rm Wilson}_{d_3} 
&=&\Bigg[ 1 -\frac{g_0^2}{16\pi^2} C_F \Bigg(\frac{35}{18} \log a^2 \mu^2 
-29.749107  +\frac{5}{6}(1-\alpha) \Bigg) \Bigg] O_{d_3}^{\rm tree} 
\nonumber \\
&& \qquad \qquad - 2.881820 \, \frac{1}{a} \, \frac{g_0^2}{16\pi^2} C_F \Big( 
  \bar{\psi} \gamma_4 \gamma_5 \gamma_{\{1} D_2 D_{3\}} \psi  
 -\bar{\psi} \gamma_1 \gamma_5 \gamma_{\{4} D_2 D_{3\}} \psi \Big) \nonumber \\
\widehat{O}^{\rm Wilson}_{t_1} 
&=&\Bigg[ 1 -\frac{g_0^2}{16\pi^2} C_F \Bigg(\log a^2 \mu^2 
+16.018079 +(1-\alpha) \Bigg) \Bigg] O_{t_1}^{\rm tree} 
\nonumber \\
\widehat{O}^{\rm Wilson}_{t_2} 
&=&\Bigg[ 1 -\frac{g_0^2}{16\pi^2} C_F \Bigg(3 \log a^2 \mu^2 
-4.477540 +\frac{3}{2}(1-\alpha) \Bigg) \Bigg] O_{t_2}^{\rm tree} 
\nonumber \\
\widehat{O}^{\rm Wilson}_{t_3} 
&=&\Bigg[ 1 -\frac{g_0^2}{16\pi^2} C_F \Bigg(\frac{13}{3} \log a^2 \mu^2 
-19.757642  +\frac{11}{6}(1-\alpha) \Bigg) \Bigg] O_{t_3}^{\rm tree} 
. \nonumber
\end{eqnarray}
We can see that for Wilson fermions the operators $O_{d_1}$, $O_{d_2}$ and 
$O_{d_3}$ are not multiplicatively renormalized. Each of them mixes with 
an operator which is one dimension lower~\cite{gea}, and the corresponding 
mixing coefficients, which seem to be gauge-invariant, are proportional to 
$r/a$. Thus, these mixings would be zero for naive fermions, and they are 
akin to the $\Sigma_0$ term in the Wilson quark self-energy, which is 
responsible for the additive renormalization of quark masses when they are 
not anymore protected by chirality. In fact, the lower-dimensional operators 
above are all mass terms of the form~(\ref{eq:massope}) coming from the OPE 
expansions in DIS. All this hints to a connection with chirality in the 
mixings of the $O_{d_n}$ operators too, and we have indeed shown in 
Eq.~(\ref{eq:ovres}) that in the calculations done with overlap fermions, 
where chirality is conserved, there is no trace of this kind of terms 
and these operators are then multiplicatively renormalized.

We can also notice from the Wilson results above that another consequence 
of the breaking of chiral invariance is that the renormalization constants 
of $O_{v_4,d}$ and $O_{a_4,d}$ are not equal, and the same happens for 
$O_{v_4,e}$ and $O_{a_4,e}$.

The perturbative calculations with Wilson fermions of the renormalization 
constants for the operators $O_{v_4,d}$, $O_{a_4,d}$, $O_{v_4,e}$, $O_{d_2}$ 
and $O_{t_1}$ have been already done in the past in Feynman gauge. We agree 
with the results for $O_{v_4,d}$ and $O_{a_4,d}$ in Ref.~\cite{bhnps}, 
but we find discrepancies with the renormalization of $O_{v_4,e}$ in 
Ref.~\cite{gea}~\footnote{In this case we do not even agree on the results 
for the individual proper diagrams, which are given in Table 5 of 
Ref.~\cite{gea}. In that Table, the operator tadpole is given the value 
that would be appropriate for $O_{v_4,d}$, but this cannot be true for
$O_{v_4,e}$, since there are additional Wick contractions between the $
A_\mu$s originating from the equal indices.}. We agree also with the results 
for $O_{d_2}$ in Ref.~\cite{gea} and for $O_{t_1}$ in Ref.~\cite{calcoli2}.

The connection of Wilson fermions with the continuum 
$\overline{\rm{MS}}$ is given by
\begin{eqnarray}
\widehat{O}^{\overline{\rm{MS}}}_{v_4,d} 
&=&\Bigg[ 1 -\frac{g_0^2}{16\pi^2} C_F \Bigg(\frac{157}{30} \log a^2 \mu^2 
-23.357525 \Bigg) \Bigg] O^{\rm lat~(Wilson)}_{v_4,d} \nonumber \\ 
\widehat{O}^{\overline{\rm{MS}}}_{a_4,d} 
&=&\Bigg[ 1 -\frac{g_0^2}{16\pi^2} C_F \Bigg(\frac{157}{30} \log a^2 \mu^2 
-23.733869 \Bigg) \Bigg] O^{\rm lat~(Wilson)}_{a_4,d} \nonumber \\  
\widehat{O}^{\overline{\rm{MS}}}_{v_4,e} 
&=&\Bigg[ 1 -\frac{g_0^2}{16\pi^2} C_F \Bigg(\frac{157}{30} \log a^2 \mu^2 
-25.503033 \Bigg) \Bigg] O^{\rm lat~(Wilson)}_{v_4,e}
\label{eq:wilsontomsbar}\\ 
\widehat{O}^{\overline{\rm{MS}}}_{a_4,e} 
&=&\Bigg[ 1 -\frac{g_0^2}{16\pi^2} C_F \Bigg(\frac{157}{30} \log a^2 \mu^2 
-25.989604 \Bigg) \Bigg] O^{\rm lat~(Wilson)}_{a_4,e} \nonumber \\ 
\widehat{O}^{\overline{\rm{MS}}}_{t_1} 
&=&\Bigg[ 1 -\frac{g_0^2}{16\pi^2} C_F \Bigg(\log a^2 \mu^2 
+17.018079 \Bigg) \Bigg] O^{\rm lat~(Wilson)}_{t_1} \nonumber \\ 
\widehat{O}^{\overline{\rm{MS}}}_{t_2} 
&=&\Bigg[ 1 -\frac{g_0^2}{16\pi^2} C_F \Bigg(3 \log a^2 \mu^2  
 +0.522460 \Bigg) \Bigg] O^{\rm lat~(Wilson)}_{t_2} \nonumber \\ 
\widehat{O}^{\overline{\rm{MS}}}_{t_3} 
&=&\Bigg[ 1 -\frac{g_0^2}{16\pi^2} C_F \Bigg(\frac{13}{3} \log a^2 \mu^2 
-11.868753 \Bigg) \Bigg] O^{\rm lat~(Wilson)}_{t_3} , \nonumber 
\end{eqnarray}
and the numbers for $\beta=6.0$, $\mu=1/a$ and $N_c=3$ are given in 
Eq.~(\ref{eq:overlapwilsonmsbar}).
We have not included here $O_{d_1}$, $O_{d_2}$ and $O_{d_3}$,
as their renormalization involves mixing coefficients that diverge in the
limit of zero lattice spacing, and as such it is of no use to compute
the connection to $\overline{\rm{MS}}$ for these operators in perturbation
theory with Wilson fermions.

\section{Conclusions}

Overlap fermions are one of the most promising formulations for putting
chiral fermions on the lattice and for studying long-standing problems
linked with chirality. In this paper we have computed the renormalization 
constants of the lowest moments of various structure functions which give 
a complete description of the quark momentum and spin at leading twist. 
We have computed these constants also with Wilson fermions, since the 
renormalization of many of these operators had never before been computed 
on the lattice. Chiral symmetry plays an important role in the structure 
of the strong radiative corrections. In particular, the overlap is the only
case in which it has yet been demonstrated that all the operators we have 
considered are multiplicatively renormalized.

The numbers presented here are also valid in the unquenched case if one 
deals with flavor non-singlet quark operators, which do not mix with gluon 
operators and for which internal quark loops never have the chance to come 
to play at 1-loop level. However, the numbers for the transversity operators 
can be considered unquenched also for flavor singlet quark operators, since 
there are no gluon operators with the same quantum numbers.

We remind again that all the renormalization constants presented in this 
paper are already fully $O(a)$ improved in the overlap case. In the Wilson 
case removing all order $a$ effects would instead involve a great amount 
of additional calculations.

With overlap fermions 1-loop corrections are substantially larger than 
for Wilson fermions, with the primary contribution arising from quark 
self-energies. This is an important physical effect that should be understood, 
and may ultimately suggest an appropriate form of resummation or tadpole
improvement.

\section*{Acknowledgment}
I would like to thank John W. Negele for critically reading the manuscript
and discussion.
This work has been supported in part by the U.S. Department of Energy (DOE) 
under cooperative research agreement DE-FC02-94ER40818.
Both the FORM and Fortran computations have been done at MIT on a few
Pentium III PCs running on Linux.

\appendix

\section{Quark self-energy}

\begin{table}[hbtp]
\begin{center}
\vspace{0.5cm}
  \begin{tabular}[btp]{|r||r|r|r|} \hline
$\rho$ & non-tadpole self-energy  & leg tadpole & total self-energy  \\ 
\hline
0.2 &  -27.511695 +11.911596 $\xi$  &  -213.087934 -7.119586 $\xi$  & 
      -240.599629 +4.792010 $\xi$  \\
0.3 &  -23.687573 +11.098129 $\xi$  &  -131.723110 -6.306119 $\xi$  &  
      -155.410693 +4.792010 $\xi$  \\
0.4 &  -21.172454 +10.520210 $\xi$  &   -91.817537 -5.728200 $\xi$  &  
      -112.989991 +4.792010 $\xi$  \\
0.5 &  -19.337313 +10.071356 $\xi$  &   -68.315503 -5.279346 $\xi$  &   
       -87.652816 +4.792010 $\xi$  \\
0.6 &  -17.912921  +9.704142 $\xi$  &   -52.931363 -4.912132 $\xi$  &   
       -70.844284 +4.792010 $\xi$  \\
0.7 &  -16.760616  +9.393275 $\xi$  &   -42.140608 -4.601265 $\xi$  &   
       -58.901224 +4.792010 $\xi$  \\
0.8 &  -15.800204  +9.123666 $\xi$  &   -34.193597 -4.331656 $\xi$  &   
       -49.993801 +4.792010 $\xi$  \\
0.9 &  -14.981431  +8.885590 $\xi$  &   -28.125054 -4.093580 $\xi$  &   
       -43.106485 +4.792010 $\xi$  \\
1.0 &  -14.270881  +8.672419 $\xi$  &   -23.359746 -3.880409 $\xi$  &   
       -37.630627 +4.792010 $\xi$  \\
1.1 &  -13.645294  +8.479438 $\xi$  &   -19.534056 -3.687428 $\xi$  &   
       -33.179350 +4.792010 $\xi$  \\
1.2 &  -13.087876  +8.303183 $\xi$  &   -16.407174 -3.511173 $\xi$  &   
       -29.495050 +4.792010 $\xi$  \\
1.3 &  -12.586126  +8.141044 $\xi$  &   -13.813486 -3.349034 $\xi$  &   
       -26.399612 +4.792010 $\xi$  \\
1.4 &  -12.130497  +7.991018 $\xi$  &   -11.635482 -3.199008 $\xi$  &    
       -23.765979 +4.792010 $\xi$  \\
1.5 &  -11.713524  +7.851554 $\xi$  &    -9.787582 -3.059544 $\xi$  &   
       -21.501106 +4.792010 $\xi$  \\
1.6 &  -11.329238  +7.721442 $\xi$  &    -8.206069 -2.929432 $\xi$  &   
       -19.535307 +4.792010 $\xi$  \\
1.7 &  -10.972744  +7.599750 $\xi$  &    -6.842630 -2.807740 $\xi$  &   
       -17.815374 +4.792010 $\xi$  \\
1.8 &  -10.639905  +7.485778 $\xi$  &    -5.660084 -2.693768 $\xi$  &   
       -16.299989 +4.792010 $\xi$  \\
1.9 &  -10.327042  +7.379023 $\xi$  &    -4.629539 -2.587013 $\xi$  &   
       -14.956581 +4.792010 $\xi$  \\
\hline
  \end{tabular} 
\vspace{0.5cm}
\caption{Results for the quark self-energy with overlap fermions, 
where we have used the abbreviation $\xi=1-\alpha$. The first column 
(non-tadpole self-energy) refers to diagram e (or f) in Fig.~1, 
and the second column (leg tadpole) refers to diagram g (or h).}
\end{center}
\vspace{0.5cm}
\end{table}

In this appendix we report the results of 1-loop calculations regarding the 
contribution proportional to ${\rm i} p \!\!\! /$ of the quark self-energy, 
also known as $\Sigma_1$. This corresponds to the diagrams e (or f) 
and g (or h) in Fig. 1, and is necessary for the calculation of the 
renormalization constants of the operators considered in this work.

For overlap fermions we have that the total self-energy is given by
\begin{equation}
\Sigma_1^{\rm overlap} = \frac{g_0^2}{16\pi^2} C_F \Bigg[ \alpha \log a^2 p^2  
+S^{\ \alpha=1} +(1-\alpha) \, 4.792010 \Bigg],
\end{equation}
where the Feynman-gauge finite results $S^{\ \alpha=1}$ are given in 
the last column of Table A.1 (in which we have used the abbreviation 
$\xi=1-\alpha$), which also contains the results of the individual diagrams.
For the total self-energy (but not for the individual diagrams) the finite 
part proportional to $(1-\alpha)$ is independent of the $\rho$ parameter, 
and even of the fermion action used. In fact this number comes from 
integrals which have only gluonic propagators, and is given by 
$F_0-\gamma_E+1=4.792009568973\cdots $~\cite{cg}.

In the Wilson case, the value of the leg self-energy (including the tadpole) 
is
\begin{equation}
\Sigma_1^{\rm Wilson} = \frac{g_0^2}{16\pi^2} C_F \Bigg[ \alpha \log a^2 p^2  
+11.852404 +(1-\alpha) \, 4.792010 \Bigg].
\end{equation}
The result for the individual diagrams is the following: for diagram e (or f)
is
\begin{equation}
\frac{g_0^2}{16\pi^2} C_F \Bigg[ \alpha \log a^2 p^2  
-0.380646 +(1-\alpha) \, 7.850272 \Bigg] ,
\end{equation}
while the leg tadpole, i.e. diagram g (or h), is
\begin{equation}
g_0^2 C_F \Bigg[ \frac{1}{2} Z_0 -(1-\alpha) \frac{1}{8} Z_0 \Bigg] 
= \frac{g_0^2}{16\pi^2} C_F \Bigg[ 12.233050 -(1-\alpha) \, 3.058262 \Bigg],
\end{equation}
where $Z_0=0.154933390231\cdots $.

\section{Results of the individual diagrams}

\begin{table}[hbtp]
\begin{center}
\vspace{0.5cm}
  \begin{tabular}[btp]{|r||r|r|r|} \hline
 & vertex (diagram a)  & sails (diagrams b + c) & op. tadpole
(diagram d)  \\ 
\hline
$O_{v4,d}$ &  0.70281 -0.53019 $\xi $  &  -9.50061 -7.02363 $\xi $  & 
            -36.69915 +4.59515 $\xi $  \\
$O_{a4,d}$ &  0.70281 -0.53019 $\xi $  &  -9.50061 -7.02363 $\xi $  &  
            -36.69915 +4.59515 $\xi $  \\
$O_{v4,e}$ &  0.61867 -0.05926 $\xi $  &  -7.80424 -7.96551 $\xi $  &  
            -40.51969 +5.06609 $\xi $  \\
$O_{a4,e}$ &  0.61867 -0.05926 $\xi $  &  -7.80424 -7.96551 $\xi $  &  
            -40.51969 +5.06609 $\xi $  \\
$O_{d1}$   &  2.81906 -0.65840 $\xi $  &   5.83620 -7.19187 $\xi $  &  
            -12.23305 +3.05826 $\xi $  \\
$O_{d2}$   &  1.15709 -0.49753 $\xi $  &   2.36476 -7.87216 $\xi $  &  
            -24.46610 +4.07768 $\xi $  \\
$O_{d3}$   &  0.64918 -0.45310 $\xi $  &   0.46340 -8.10072 $\xi $  &  
            -36.69915 +4.59515 $\xi $  \\
$O_{t1}$   &  4.35437 -3.79201 $\xi $  &  0  &  0  \\
$O_{t2}$   &  1.46711 -1.23375 $\xi $  &  -5.34275 -5.11652 $\xi $  &  
            -12.23305 +3.05826 $\xi $  \\
$O_{t3}$   &  0.68662 -0.54766 $\xi $  &  -7.90490 -6.48870 $\xi $  &  
            -24.46610 +4.07768 $\xi $  \\
\hline
  \end{tabular} 
\vspace{0.5cm}
\caption{Results of the proper diagrams for overlap fermions, 
where we have used the abbreviation $\xi=1-\alpha$.}
\end{center}
\vspace{0.5cm}
\end{table}

\begin{table}[hbtp]
\begin{center}
\vspace{0.5cm}
  \begin{tabular}[btp]{|r||r|r|r|} \hline
 & vertex (diagram a)  & sails (diagrams b + c) & op. tadpole
(diagram d)  \\ 
\hline
$O_{v4,d}$ &  0.842048 -0.530195 $\xi $  &  -9.201715 -7.023629 $\xi $  &
            -36.699150 +4.595148 $\xi $  \\
$O_{a4,d}$ &  0.465704 -0.530195 $\xi $  &  -9.201715 -7.023629 $\xi $  &  
            -36.699150 +4.595148 $\xi $  \\
$O_{v4,e}$ &  0.842455 -0.059255 $\xi $  &  -7.527094 -7.965509 $\xi $  &  
            -40.519687 +5.066088 $\xi $  \\
$O_{a4,e}$ &  0.355884 -0.059255 $\xi $  &  -7.527094 -7.965509 $\xi $  &
            -40.519687 +5.066088 $\xi $  \\
$O_{d1}$   &  2.590817 -0.812814 $\xi $  &  -1.845174 -7.037458 $\xi $  & 
            -12.233050 +3.058263 $\xi $  \\
$O_{d2}$   &  0.988791 -0.585340 $\xi $  &  -4.052542 -7.784353 $\xi $  &  
            -24.466100 +4.077683 $\xi $  \\
$O_{d3}$   &  0.536160 -0.521292 $\xi $  &  -5.438520 -8.032532 $\xi $  &  
            -36.699150 +4.595148 $\xi $  \\
$O_{t1}$   &  4.165675 -3.792010 $\xi $  &  0  &  0  \\
$O_{t2}$   &  0.980373 -1.233747 $\xi $  &  -5.077267 -5.116525 $\xi $  &  
            -12.233050 +3.058263 $\xi $  \\
$O_{t3}$   &  0.465618 -0.547660 $\xi $  &  -7.609564 -6.488700 $\xi $  &  
            -24.466100 +4.077683 $\xi $  \\
\hline
  \end{tabular} 
\vspace{0.5cm}
\caption{Results of the proper diagrams for Wilson fermions, 
where we have used the abbreviation $\xi=1-\alpha$.}
\end{center}
\vspace{0.5cm}
\end{table}

\begin{table}[hbtp]
\begin{center}
\vspace{0.5cm}
  \begin{tabular}[btp]{|r||r|r|r|} \hline
 & vertex (diagram a)  & sails (diagrams b + c) & op. tadpole
(diagram d)  \\ 
\hline
$O_{d1}$   &  1.508200  &  14.735561  &  0  \\
$O_{d2}$   &  0.180285  &   4.085395  &  0  \\
$O_{d3}$   &  0.073704  &   2.808116  &  0  \\
\hline
  \end{tabular} 
\vspace{0.5cm}
\caption{Results for the mixing with the lower-dimensional operators 
for Wilson fermions.}
\end{center}
\vspace{0.5cm}
\end{table}

In this appendix we give the results for the finite parts of the proper 
diagrams: the vertex (diagram a in Fig. 1), the sails (diagrams b plus c) 
and the operator tadpole (diagram d). We have used the abbreviation 
$\xi=1-\alpha$. Table B.1 refers to overlap fermions (for $\rho=1$) and 
Table B.2 to Wilson fermions. For the latter, Table B.3 also reports 
the numbers giving the mixing with the lower dimensional operators.

Note that the sum of the $(1-\alpha )$ parts of vertex and sails is 
independent of the fermion action used, and in general this is also true 
for the individual diagrams. An exception to this are the $O_{d_n}$ operators 
and the quark self-energy, and these are also the same cases for which with 
Wilson fermions there is a 1-loop mixing with a $1/a$ divergent coefficient, 
while with overlap fermions such mixings are forbidden by chiral symmetry.

\end{document}